# Particle-In-Cell Simulation of RFQ in SSC - Linac

Xiao Chen[1,2], He Yuan[2] , Yuan You-Jin[2], Liu Yong[2], Xia Jia-Wen[2], Lu Yuan-Rong[3], Yuri Batygin[4]

[1] Graduate University of Chinese Academy of Sciences, Beijing, 100049

[2] Institute of Modern Physics, Chinese Academy of Sciences, Lanzhou, 730000

[3] Key Laboratory of Heavy Ion Physics Ministry of Education, Beijing, 100049

[4] Los Alamos National Laboratory, Los Alamos, NM, 87545

**Abstract**

A 52MHz Radio Frequency Quadrupole (RFQ) linear accelerator (linac) is designed to serve as an initial structure for the SSC-linac system (injector into Separated Sector Cyclotron). The designed injection and output energy are 3.5 keV/u and 143 keV/u, respectively. Beam dynamics study in RFQ was done using 3-dimensional particle-in-cell code BEAMPATH [1]. Simulation results show that this RFQ structure is characterized by stable value of beam transmission efficiency (at least 95%) for both zero-current mode and for space charge dominated regime. The beam accelerated in RFQ has good quality in both transversal and longitudinal directions, and could be easily accepted by Drift Tube Linac (DTL). Effects of vane errors and of the space charge on beam parameters are studied as well to define the engineering tolerance for RFQ vane machining and alignment.

# 1 Introduction

Present injector of the Heavy Ion Research Facility Lanzhou (HIRFL) contains Sector Focusing Cyclotron (SFC) together with the Separated Sector Cyclotron (SSC) to accelerate heavy ions before injection into Cooler Storage Ring (CSR) [2]. This mode is no longer able to meet the growing experimental requirements. If linear accelerator were added to HIRFL-CSR system to serve as injector for the SSC, the actual beam time available for experiments would grow up substantially during the HIRFL-CSR operation. The SSC-linac system could supply Ca, Zn, and Fe ion beams with energies from 5 MeV/u to 6 MeV/u for Super Heavy Element experiments. Beam intensity is expected to be around 1 puA. Additionally, it is suggested to inject all kind of heavy ion beams with energy of 10MeV/u into CSR. In this mode of operation, beam intensity is expected to be around 1 euA (for uranium). As a part of SSC-Linac, the 52MHz RFQ has been designed and extensively simulated by PARMTEQ-M code [7], and crosschecked by the BEAMPATH code using the particle-in cell (PIC) method.

# 2 RFQ beam dynamics

## 2.1 The specifications and choice of parameters

The RFQ structure is supposed to be used for acceleration of ion beams from 3.5 keV/u to 143 keV/u. It has been specified to have a length shorter than 3 meters and to be placed downstream to the LEBT. Design frequency was chosen to be 52 MHz. Analysis of beam dynamics shows that this value is a good compromise between the required focusing strength and the maximum achievable surface electric field. The RFQ is divided into four traditional sections as the radial matching section, the shaper, the gentle buncher, and the accelerator section. Since the injection energy is relatively low (3.5keV/u), the short electrodes are chosen to bunch the beam. This option will increase the beam energy spread in accelerator section. The bunching process results in increase of the beam density, which, in turn, increases the space charge forces, and might result in blow up of the transversal beam emittance. In order to overcome the space charge effects, the gradient of acceleration has to be enhanced rapidly. The main parameters of the RFQ are listed in Table 1, and their evolution along the structure are shown in Figure 2.

**2.2 PIC mode simulation**

Initial design of the RFQ linac is made with standard approach using the PARMTEQ-M code. Beam dynamics simulation was performed using BEAMPATH code with a beam represented as a collection of 20,000 macro-particles. The RFQ is designed to capture, bunch, and accelerate a continuous unbunched beam. However, to increase longitudinal capture efficiency of the RFQ before injection into cyclotron, a prebunching of the beam in the Low Energy Beam Transport (LEBT) line is required. The LEBT beam dynamics simulation indicate that pre-bunching of the continuous beam reduces the total phase width of the beam at the entrance of RFQ from 360 degrees to 90 degrees. The value of momentum spread of the beam in this case will be increased from 0.01% to 1%. Parameters of the initial beam are listed in Table 2. Two different versions of beam injector are compared.

Figure 3a illustrates continuous beam injection into RFQ with beam momentum spread of 0.01%. Figure 3b demonstrates particle distribution with $90^{o}$ phase width and 1% momentum spread in bunched beam injection mode. Figure 3c shows the final particle distribution under continues beam injection. The value of transmission efficiency of 93.6% is achieved in this mode of operation where beam losses are mostly concentrated in the gentle buncher section of the RFQ. Figure 3d illustrates particle distribution at the exit of RFQ at the bunched beam injection mode. In the latest case, the value of the beam transmission efficiency is close to 100%, the phase width of the extracted beam from RFQ is 7.5 degrees, and the final momentum spread is about 0.7%. The quality of the beam in the longitudinal phase space in this case is substantially better than that under the continuous beam injection.

Additional beam dynamics simulation was done using BEAMPATH code to optimize the RFQ with respect to transmission, beam quality and emittance growth. Figure 4a illustrates variation of optimized beam parameters along RFQ. Figure 4b demonstrates initial and final transverse particle distributions in RFQ. Let us note, that the value of initial beam emittance of 0.1 $\pi$ cm mrad was taken larger than that of the ECR ion source (0.06 $\pi$ cm mrad). This gives us a safety margin to cope with misalignments and the mismatches of the ECR ion source.

## 3 Effect of beam mismatch on transmission efficiency and emittance growth in RFQ

Unmatched conditions for the injected beam result in additional oscillations of the beam in the focusing structure. The mismatch errors are the reason for decrease of beam transmission efficiency and of the beam emittance growth. In the process of transversal parameters sweep, the phase width of the beam of 90 degrees and momentum spread of 1% were fixed. The values of beam envelope $R_x = \sqrt{\beta_x \varepsilon_x}$ and of the tilt of the envelope $dR_x / dz = -\alpha_x \sqrt{\varepsilon_x / \beta_x}$ varied in the vicinity of the matching points. The same procedure was done in the vertical direction. The transmission efficiency and the beam emittance growth were calculated as functions of initial beam conditions in order to test the tolerance of the mismatch errors.

Figure 5a shows dependence of the value of transmission efficiency on variation of initial beam parameters. Transmission efficiency reaches the value of 95% and more, while $R_x$ changes from 0.25 cm to 0.55 cm and $dR_x / dz$ varies from 0 rad to -0.185 rad. Figures 5b and 5c illustrate final RFQ beam emittance due to variations of initial beam parameters. The study indicates that if the initial values of RX and DRXDZ vary in the area with high transmission efficiency, no significant beam emittance growth is observed. When initial values of $R_x$ and $dR_x / dz$ are outside of this region, serious emittance growth takes place.

## 4 Decrease of transmission efficiency due to RFQ vane manufacturing errors and beam space charge

Random errors in manufacturing of RFQ vane tips result in growth of amplitude of transversal and longitudinal particle oscillations [5]. Analytical treatment shows monotonous enlargement of transversal oscillation amplitude $r_{max}$ and vertical size of separatrix $<\Delta g>=<\Delta P>/v_s$ after passing through the RFQ section with $N$ cells:

$$<\delta r_{\max}>^2 = 2N[<\delta r_o>^2 + r_{\max}^2(<\frac{\delta U}{U}>^2 + 4<\frac{\delta R_o}{R_o}>^2)], \quad (1)$$

$$<\Delta g>^2 = \pi^2(\frac{\Omega_f}{\omega})^2(\frac{W_o}{\Delta W}+N)$$

$$\{(\frac{\Omega_f}{\omega})^2 ctg^2\phi_n(1-\sqrt{\frac{W_o}{W_f}})(<\frac{\delta A}{A}>^2 + <\frac{\delta U}{U}>^2) + \frac{1}{3}[1-(\frac{W_o}{W_f})^{\frac{3}{2}}]<\frac{\delta L}{L}>^2\}, \quad (2)$$

where $\delta r_o$ is the axis displacement, $\delta R_o$ is the error in average radius of the structure, $\delta L$ is the error in cell length, $\delta U$ is the inter-vane voltage instability, $\Omega_f$ is the longitudinal oscillation frequency, $\omega$ is the RF frequency, $W_o$ and $W_f$ are initial and final particle energy, $A$ is the acceleration efficiency, and $\phi$ is the synchronous phase. To study this effect via BEAMPATH simulation, the following parameters were randomly distributed at every cell within the max error of $\pm\delta$: cell length $L_i$, aperture $a_i$, maximum distance from the axis to electrodes $ma_i$, and axis displacement $\delta r_{oi}$.

According to the simulation (see Table3), the error of 50 microns does not create any serious degradation of the beam parameters while error of 100 microns could cause notable decrease of beam transmission efficiency. During the manufactory process, the engineering tolerance of 50 microns is thus adopted for vane tips fabrications. Trajectories were calculated using 100 steps per cell in the field, which is combination of RFQ, potential and self-field of the train of bunches. Space charge field was found at each time step from solution of the 3-dimensional Poisson's equation on the mesh with the Dirichlet boundary conditions for potential on the surface of a square aperture, and periodic conditions in longitudinal direction.

Operation of RIKEN RFQ linac under similar conditions indicated that 90% of transmission efficiency is obtained steadily following this simulation [6].

## 5 Conclusions

Simulation of the RFQ dynamics via BEAMPATH code is similar to the results of the PARMTEQ-M code and the beam from the RFQ could be accepted smoothly by DTL linac. The parameters sweep analysis is done to test the RFQ acceptances in both transversal and longitudinal directions. The vanes manufacturing errors and the space charge effect have been studied to confirm the engineering tolerance. The beam transmission efficiency and the emittance growth are not sensitive to beam current when the current is lower than 2 mA and the beam dynamics are not strongly influenced by the space charge effects.

Table 1. Specification for the RFQ

| | |
|---|---|
| Particle | $U^{34+}$ |
| Input Energy | 3.5 keV/u |
| Output Energy | 143 keV/u |
| RF Frequency | 52 MHz |
| Max Current | 0.5 mA |
| Max Surface field | 13.4 MV/m |
| RFQ Length | 251.01 cm |
| Intervane Voltage | 68 kV |
| Max Modulation | 1.966 |
| Synchronous Phase Final | -27° |

Table 2. Initial beam parameters.

| Parameters | Value | Unit |
|---|---|---|
| Input Energy | 3.5 | kev/u |
| Emittance | 0.1 | $\pi$.cm.mrad |
| Distribution | Water-bag | |
| $\alpha_x$ and $\alpha_y$ | 1.02 | |
| $\beta_x$ and $\beta_y$ | 4.24 | cm/rad |
| Momentum Spread | 0.01 or 1 | % |
| Phase width | 360 or 90 | degrees |

Table 3. Beam transmission efficiency as a function of errors in vane tip fabrication and beam space charge.

| | $\delta$ (microns) | I=0mA | I=1mA | I=2mA |
|---|---|---|---|---|
| 1 | 0 | 1.0000 | 0.9995 | 0.9760 |
| 2 | 50 | 0.9810 | 0.9975 | 0.9520 |
| 2 | 75 | 0.8300 | 0.8490 | 0.8415 |
| 3 | 100 | 0.6035 | 0.6700 | 0.6920 |

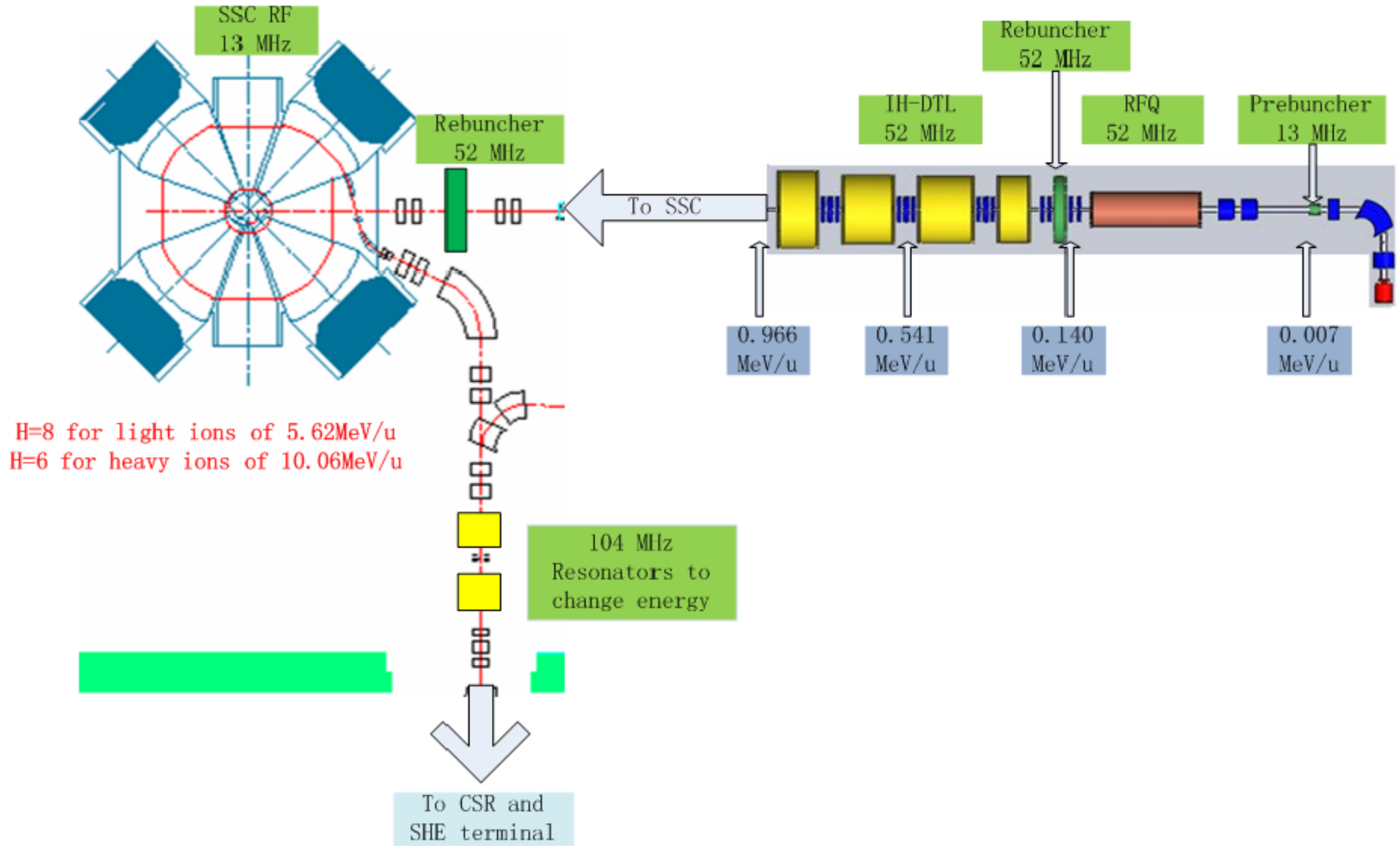

Fig. 1. Conceptual design of SSC-linac.

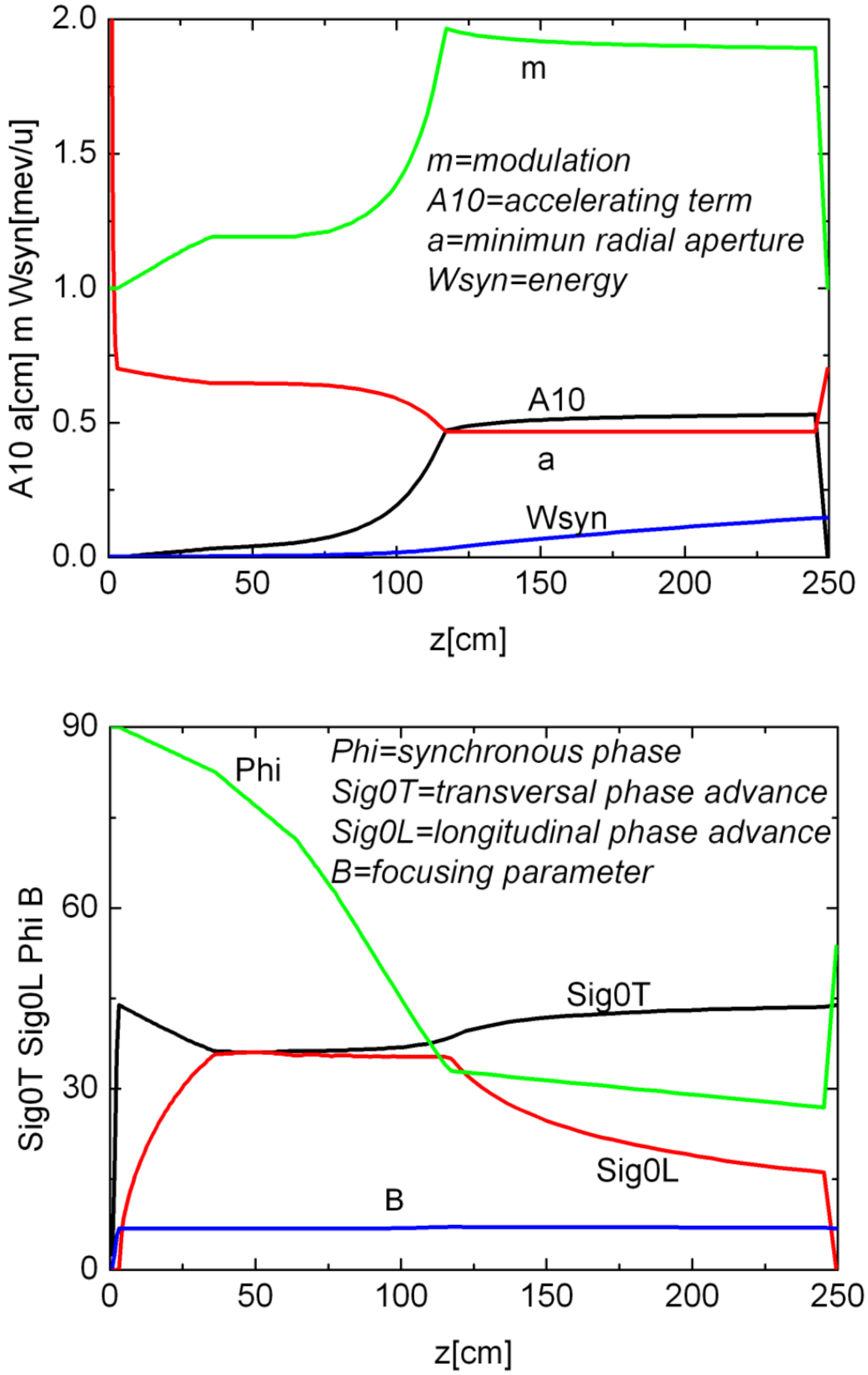

Fig. 2. Optimized design parameters of the RFQ.

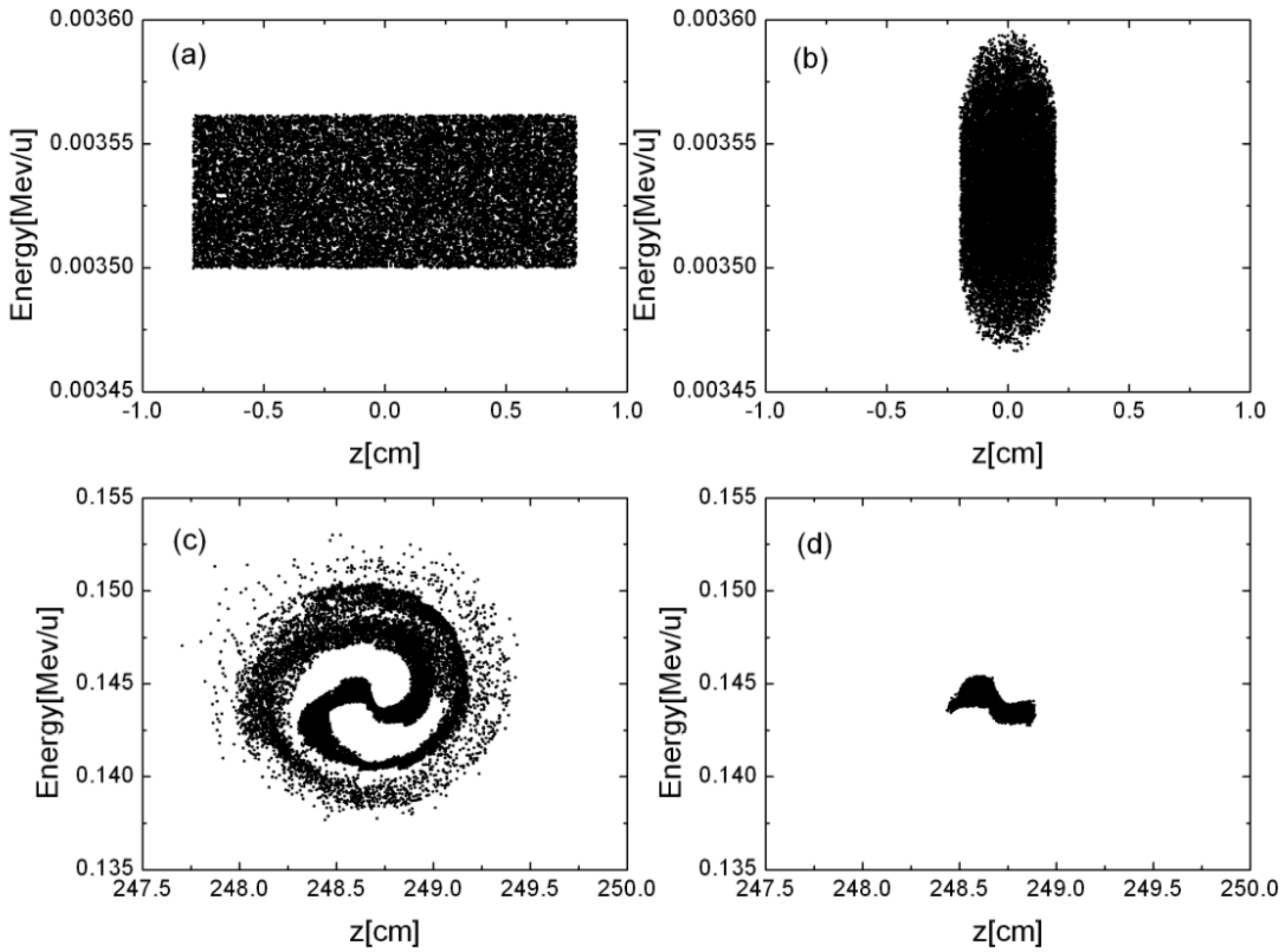

Fig. 3. (Left) particle distribution in z-energy space at the entrance and at the exit of RFQ for the continuous injected beam and (right) for the bunched beam injection.

a)

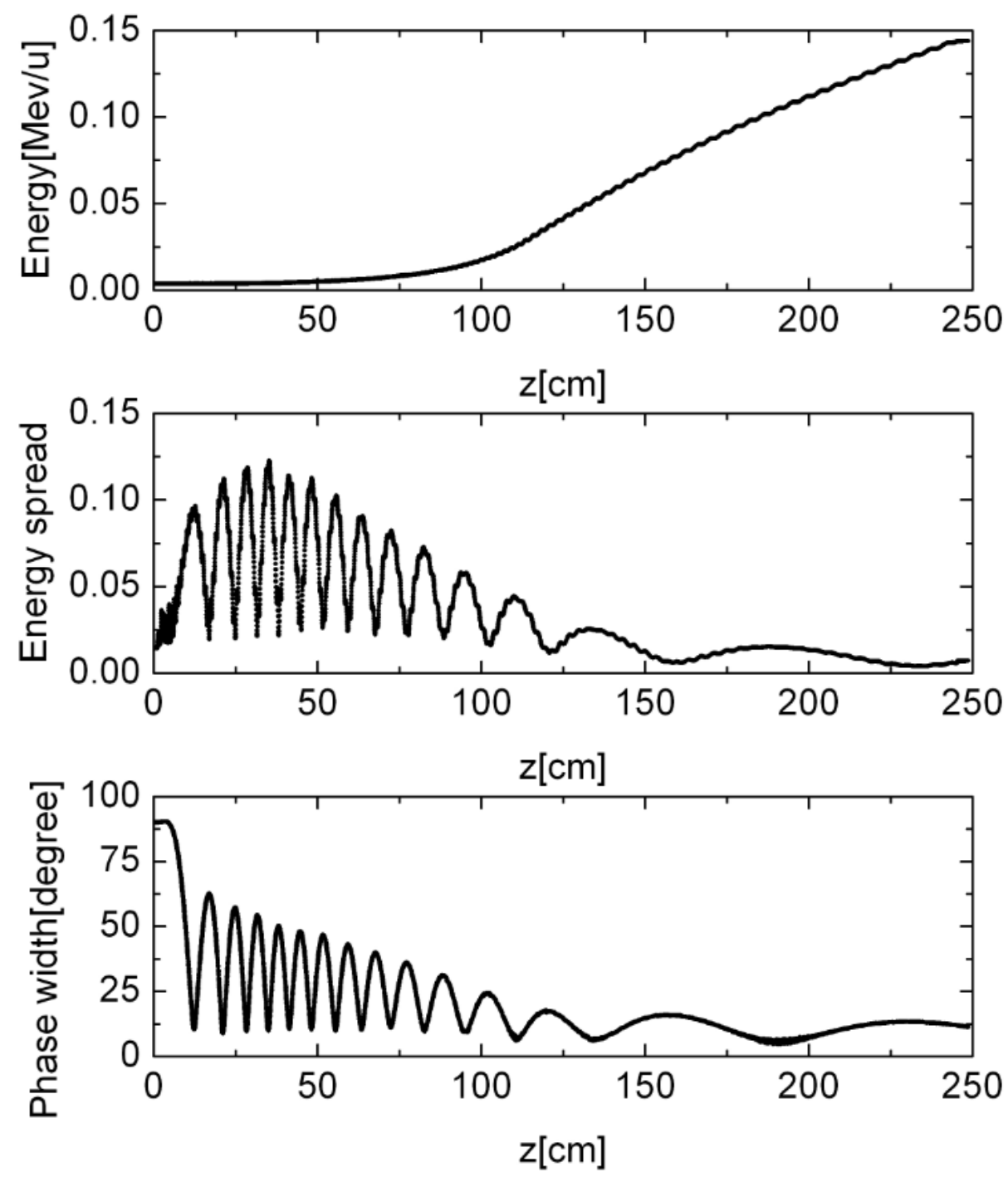

b)

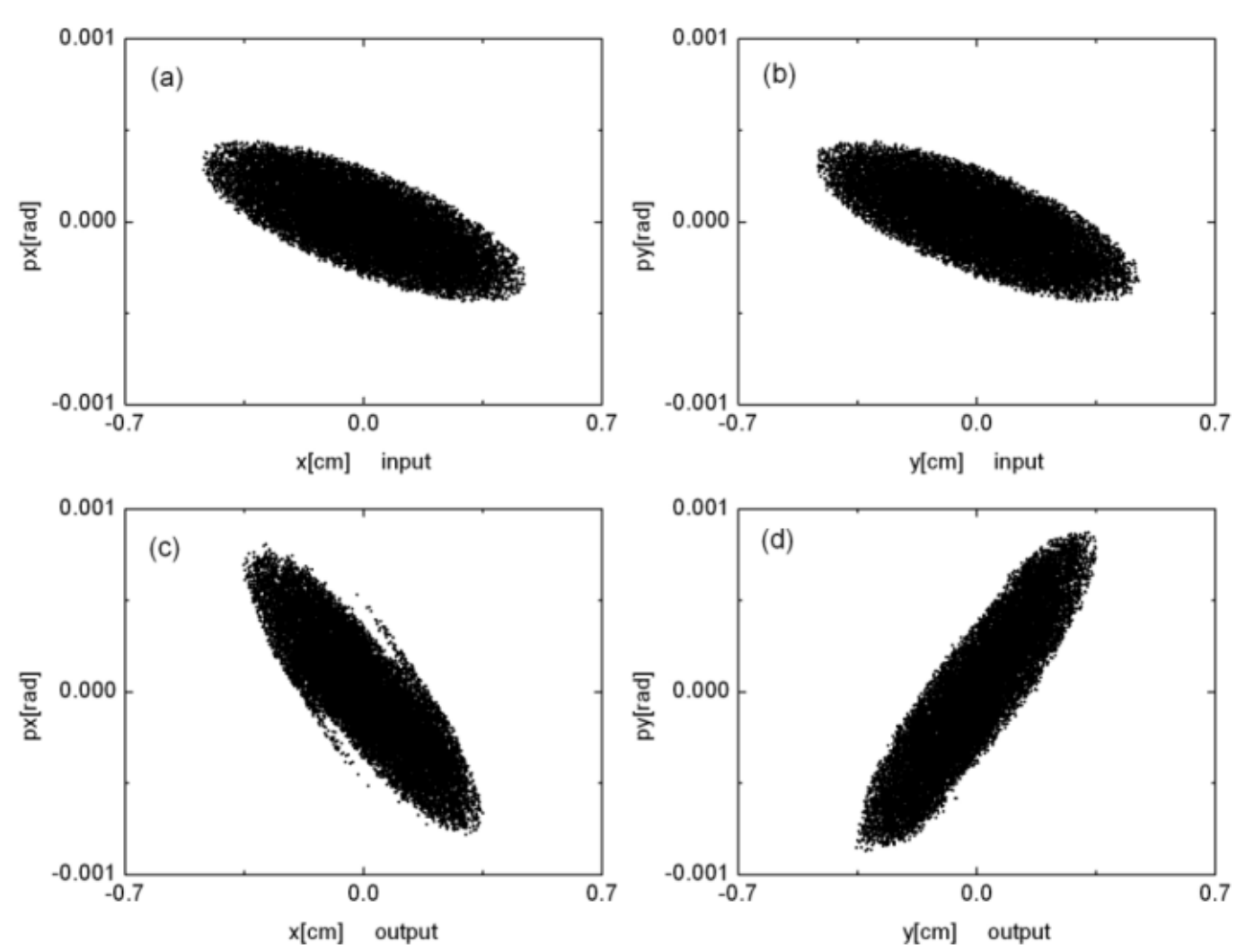

Fig. 4. (a) Energy increase, energy spread , and variation of phase width along the RFQ, (b) initial and final phase space plots in the horizontal and vertical directions.

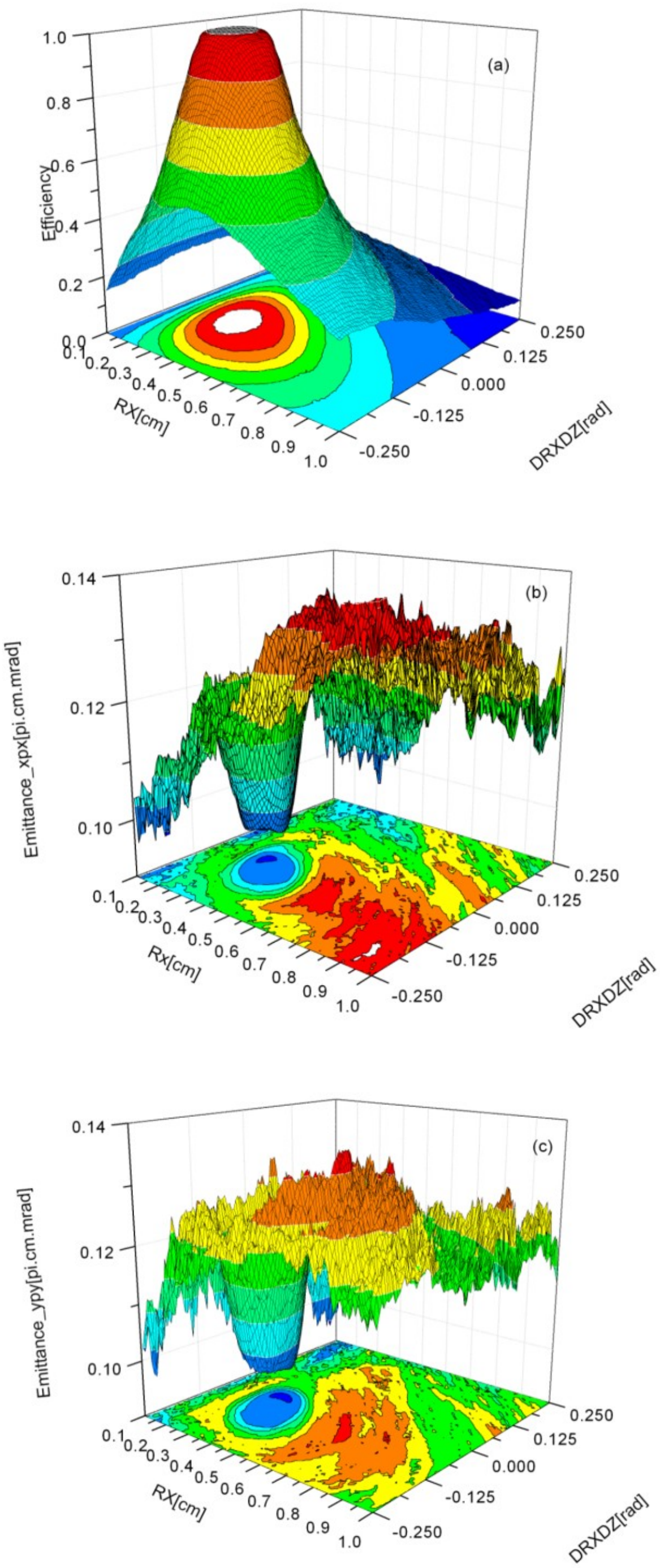

Fig. 5. Decrease of transmission efficiency and beam emittance growth due to mismatch of the beam with RFQ.

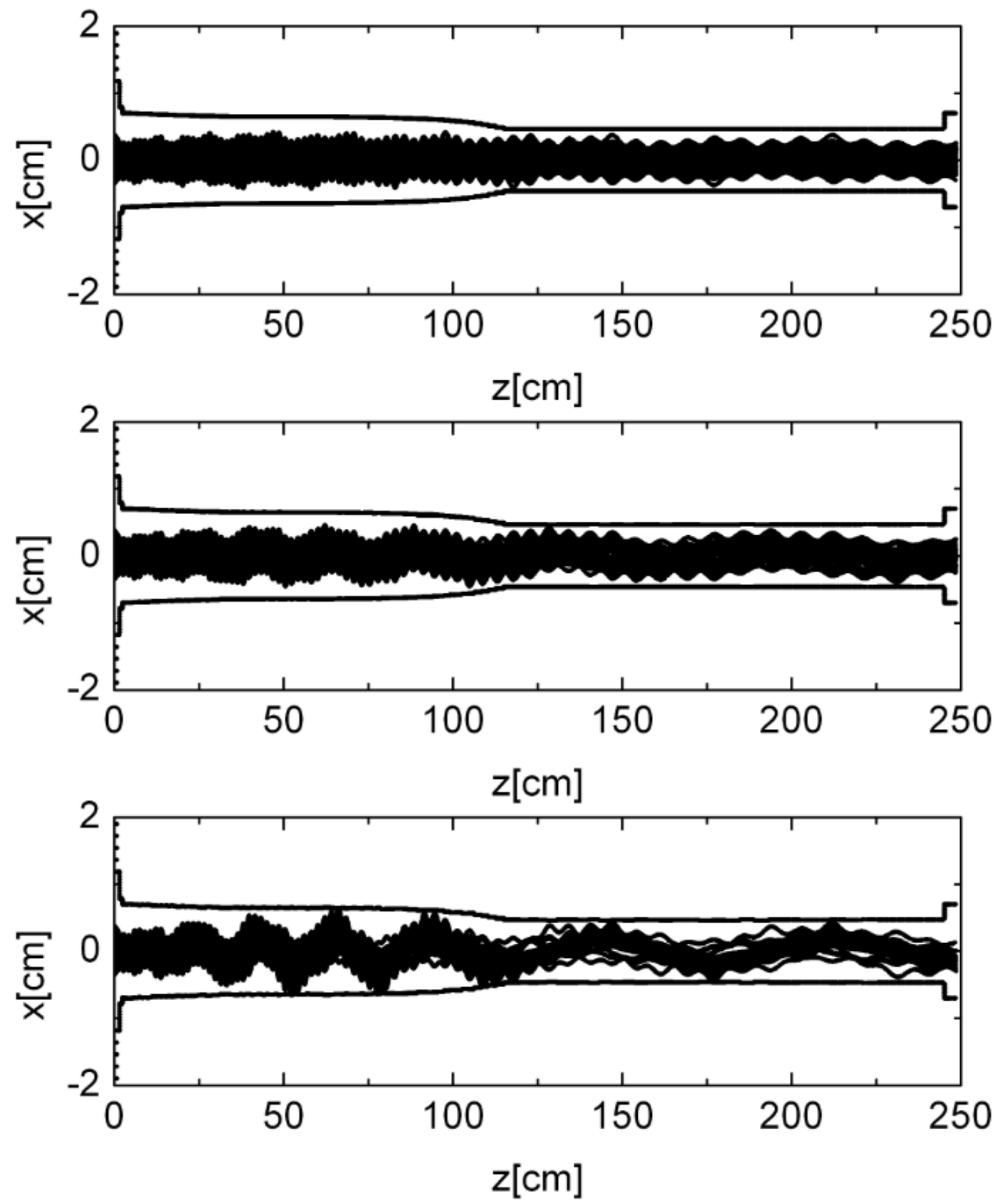

Fig. 6. Particle trajectories in RFQ with manufacturing errors of 0 um, 50 um, and 100 um at beam current of 0 mA.